\documentstyle[11pt,newpasp,twoside,psfig]{article}
\begin{document}

\title{A search for LSB dwarf galaxies in various environments}
\author{Sarah Roberts, Jonathan Davies, Sabina Sabatini}
\affil{Cardiff University, Department of Physics and Astronomy, 5, The Parade, Newport Road,
 Cardiff, UK, CF24 3YB}

\begin{abstract}
The varying dwarf galaxy populations in different environments poses a problem for Cold Dark Matter (CDM) hierarchical clustering models. In this paper we present results from a survey conducted in different environments to search for low surface brightness (LSB) dwarf galaxies. 
\end{abstract}
\vspace{-1cm}
\section{Introduction}
According to standard Cold Dark Matter (CDM) hierarchical clustering theory, there should be
 numerous low mass dark matter halos present in the Universe today. If these halos contain
sufficient stars, they should be detectable as dwarf galaxies. Observationally this appears to 
be true for clusters of galaxies where the galactic density is high, but not so for the lower
 density environments. We conducted a search for these objects in the Millennium galaxy strip which runs
 along the celestial equator in the field, passing through filaments and voids. It is
 therefore an excellent data set for studies into the influence of the environment on dwarf galaxy 
populations. We compare these results with those from similar surveys carried out in the Virgo
 and Ursa Major (UMa) clusters. Our results are unique as the three surveys were conducted using the
 same instrument, same technique (exposure time, filter band) and same selection criteria, thus
 we can be sure that we are comparing 'like with like'. 

Low surface brightness (LSB) galaxies are difficult to detect as their surface brightnesses are 
below that of the sky ($\geq 23 mag/arcsec^{2}$). The detection algorithm that we developed for this 
project is optimised for the detection of faint, diffuse objects on CCD frames (see Sabatini et al. 2003).
 To ensure that the objects picked out by the algorithm are 
actually dwarf galaxies and not background contamination, a selection criteria based on morphology and 
magnitude is applied to the objects. This criteria was chosen following simulations of a cone 
of the universe randomly populated with galaxies, as detailed in Roberts et al (2003). 
\section{Results and Discussion}
\begin{table}
{\small
\caption{Summary of results for the three surveys, compared with predictions from CDM}
\begin{tabular}{||c|c|c||} \hline
{\em{Environment}} & {\em{Description}} & {\em{DGR}\footnotemark[1]}\\ \hline

MGS & Passes through regions of high and low density & 6:1 \\
UMa & Low density cluster & - \\
Virgo & High density cluster & 20:1 \\
2dF LF ($\alpha$ =-1.2) & 2dF survey results (Norberg et al. 2002) & 7:1  \\
CDM ($\alpha$ =-1.6) & Schechter LF integration & 370:1 \\
CDM ($\alpha$ =-2.0) & Schechter LF integration & 8500:1 \\
\hline
\end{tabular}}
\end{table}

\footnotetext [1] {We define a dwarf to giant ratio (DGR) as the number of dwarfs with -10$\geq M_{B}\geq$-14 divided by the number of giants with $M_{B}\leq$-19.}

We have presented the results obtained for 3 surveys carried out in very different environments (table 1).
  We find 
a DGR in the field of 6:1, compared to a value of 20:1 in the Virgo cluster. This very large ratio of 
dwarf to giant galaxies found in the Virgo cluster indicates that this region is very different to lower
 density clusters such as UMa, and the field where we find relatively few dwarfs.

Our results for the DGR of the MGS are consistent with those derived from the recent redshift survey determinations of the field LF made by 2dF (Norberg et al. 2002) even though we sample to some two magnitudes fainter in central surface brightness and magnitude. There is no hidden population of dwarf galaxies that have been missed by the redshift surveys.

These observational results are in disagreement with most predictions made by CDM models, commonly referred to as the 'substructure' problem. The models predict far more small dark matter halos than observations detect (Kauffman et al. 1993). A number of theories have been put forward to explain the apparent difference in the observed number of dwarf galaxies in different environments. These range from those which try and explain how more dwarfs may form in rich environments, such as galaxy harassment (Moore et al 1996) or the external pressure confining the ejected galaxy gas (Babul \& Rees 1992), to those ideas which emphasise the suppression of dwarf galaxy formation in the field such as supernovae wind expulsion and galaxy squelching (Tully et al 2002). 
Detailed observations of dwarf galaxies provide a challenge to the 'concordance' cosmological model to which CDM is central. Dwarf galaxies are found in large numbers in rich clusters, but not in less dense galactic environments. For the CDM model to remain viable it has to provide a satisfactory solution to this problem. At present, it is not clear which, if any, of the mechanisms described above would be the best to help provide this solution.

\end{document}